# NEUTRONIC CHARACTERIZATION OF THE MEGAPIE TARGET

Stefano Panebianco[1, a], Olivier Bringer[1], Pavel Bokov[1], Sébastien Chabod[1], Frédéric Chartier[2], Emmeric Dupont[1], Diane Doré[1], Xavier Ledoux[3], Alain Letourneau[1], Ludovic Oriol[4], Aurelien Prevost[1], Danas Ridikas[1], Jean-Christian Toussaint[1]

[1] CEA Saclay, DSM/DAPNIA, F-91191 Gif-sur-Yvette, France
[2] CEA Saclay, DEN/DPC, F-91191 Gif-sur-Yvette, France
[3] CEA/DAM Ile-de-France, DPTA/SPN, F-91680 Bruyères-le-Châtel, France
[4] CEA Cadarache, DEN/DER, F-13108 Saint-Paul-lez-Durance, France

*The MEGAPIE project is one of the key experiments towards the feasibility of Accelerator Driven Systems. On-line operation and post-irradiation analysis will provide the scientific community with unique data on the behavior of a liquid spallation target under realistic irradiation conditions. A good neutronics performance of such a target is of primary importance towards an intense neutron source, where an extended liquid metal loop requires some dedicated verifications related to the delayed neutron activity of the irradiated PbBi. In this paper we report on the experimental characterization of the MEGAPIE neutronics in terms of the prompt neutron (PN) flux inside the target and the delayed neutron (DN) flux on the top of it. For the PN measurements, a complex detector, made of 8 microscopic fission chambers, has been built and installed in the central part of the target to measure the absolute neutron flux and its spatial distribution. Moreover, integral information on the neutron energy distribution as a function of the position along the beam axis could be extracted, providing integral constraints on the neutron production models implemented in transport codes such as MCNPX. For the DN measurement, we used a standard $^3$He counter and we acquired data during the start-up phase of the target irradiation in order to take sufficient statistics at variable beam power. Experimental results obtained on the PN flux characteristics and their comparison with MCNPX simulations are presented, together with a preliminary analysis of the DN decay time spectrum.*

## I. INTRODUCTION

### I.A. Motivations

Based on the initiative of six European institutions (PSI, FZK, CEA, SCK-CEN, ENEA, CNRS), JAEA (Japan), DOE (US) and KAERI (Korea), the MEGAwatt PIlot Experiment (MEGAPIE) started officially in the year 2000 aiming to design, build and safely operate a liquid metal (Lead-Bismuth Eutectic, LBE) spallation target at 1 MW beam power[1]. The MEGAPIE target was delivered to PSI in 2005, installed and tested in the SINQ hall during spring 2006 and successfully irradiated during 4 months starting in mid-August.

It is considered as an essential step towards the development of high power spallation targets to produce intense neutron sources, neutrino beams and a fundamental technological piece for RIB (Radioactive Ion Beam) facilities and nuclear waste incinerators driven by accelerators (ADS). In particular, it is meant to be the key experiment for high power window targets based on heavy liquid metal technology. In addition, it will also contribute to the lead-cooled reactor research within the GEN-IV initiative.

The increased proton beam power needed by new generation targets results in heat deposition constraints leading to innovative designs based on heavy liquid metal (HLM). Present spallation targets aiming at beam power of 1 MW or higher (e.g., MEGAPIE (PSI), SNS (US), JSNS (Japan), ESS and EURISOL (both Europe)), all focus on liquid metal targets. In fact, the use of liquid metal loop can solve some difficult problems for high-power spallation sources, mainly related to the evacuation of the deposited heat. On the other hand, it also introduces some new issues that must be addressed (e.g., corrosion, resistance of the target window, leakage during operation, etc.).

Although spallation physics models are nowadays reliable and well qualified against a lot of experimental nuclear data, a spallation target such as MEGAPIE is a complex system for which an accurate simulation of the neutron flux is fundamental. Furthermore, a precise neutronic characterisation is crucial for future ADS developments in order to address the possibility to transmute minor actinides in such a system. These are the reasons why we have designed and built a neutron detector to measure "in situ" and to characterise the inner neutron flux of the target under irradiation. Coupled with very detailed Monte Carlo simulations, these integral measurements should provide the scientific community with accurate data on the neutron generation of such a system to constrain commonly used neutron production models including neutron transport phenomena. Moreover, other effects as the influence of spallation residues accumulation on the neutron balance or the temperature on the neutron energy spectrum could be assessed.

---
[a] E-mail: stefano.panebianco@cea.fr

Finally, in the framework of the R&D on future spallation sources, neutrino factories or RIB facilities, radioprotection issues are clearly of major concern. Indeed, radioactive nuclides produced in liquid metal targets are transported into hot cells, pumps, or close to electronics with radiation sensitive components. Besides the considerable amount of gamma decay activity in the irradiated liquid metal, a significant amount of the Delayed Neutron (DN) precursor activity can be accumulated in the target fluid. The transit time from the front of a liquid metal target into areas where DNs may be important, can be as short as a few seconds, i.e. well within one half-life of many DN precursors. Therefore, it is very important to evaluate the DN flux as a function of position and determine if DNs may contribute significantly to the activation and dose rates.

### I.B. The MEGAPIE target

The target, installed in the SINQ location at the Paul Sherrer Institute (Switzerland), has been designed to accept a proton current of 1.74 mA. The thermal energy deposited in the lower part of the target is removed by forced convection. The LBE is driven by the main inline electromagnetic pump, then passes through a 12-pin heat exchanger (HX) and returns to the spallation region. The heat is evacuated from the heat exchanger through a diathermic oil loop to an external intermediate water cooling loop and then finally goes into the PSI existing cooling system. The beam entrance window is cooled both by the main flow and also by a cold LBE jet extracted at the heat exchanger outlet, which is pumped by a second electromagnetic pump (Fig. 1). The target has been conceived in nine sub-components, which were manufactured separately and finally assembled.

## II. DESCRIPTION OF THE EXPERIMENTS

### II.A. Prompt neutron (PN) flux measurement

The absolute inner neutron flux of the target was measured with an innovative dedicated neutron flux detector[2]. It should be noticed that a 1 MW liquid Pb-Bi spallation target like MEGAPIE constitutes a very constraining environment due to:

- *high temperature fluctuations*: around 420 °C with beam-on and 230 °C with beam-off,
- *high level of radiations*: more than $10^{13}$ n/cm$^2$/s and almost the same level of gamma rays coming from spallation reactions and activation of structural materials (as given by simulations),
- *electromagnetic perturbations* due to electromagnetic pumps.

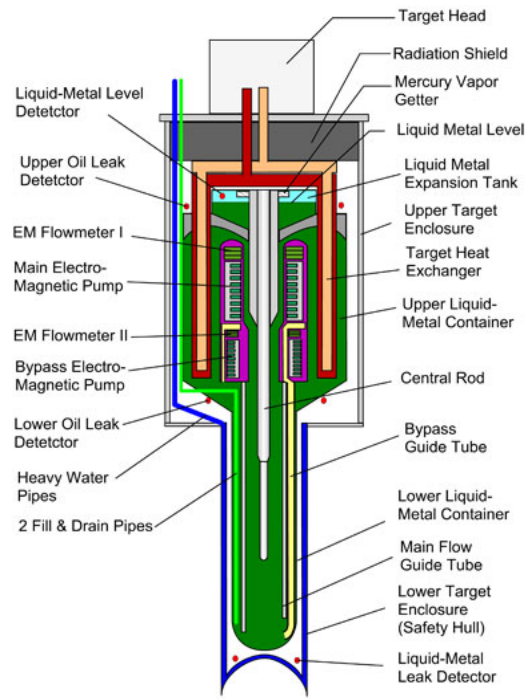

Fig. 1. Principal layout of the MEGAPIE target (not to scale).

Moreover, the overall dimensions of the central rod, where the detector was inserted, were very tiny (20 mm in diameter) and its access was impossible during the whole irradiation period.

The neutron detector, built in 2005, contains eight miniature fission-chambers already employed in the framework of the Mini-INCA project[3]. The fission chambers were adapted for the MEGAPIE specific environment, with dedicated cables, electronics and acquisition system. The entire detector is 5 m long with a 13 mm diameter in its lower part and 22 mm in the upper part. Fission chambers are located in the thinner part of the detector, to be as close as possible to the proton-beam interaction zone in the Pb-Bi (Fig. 2).

Signals from fission chambers are transported by 1 mm thick mineral cables inside the detector and connected to triaxial organic cables outside the detector to avoid electromagnetic perturbations. Fission chambers are imbedded in pairs along the axis of the detector over a 50 cm length. Each pair, except one, is made of a chamber containing $^{235}$U fissile isotope and a chamber without deposit (WD in Fig. 2) to compensate the fission signal from leakage currents or from currents induced by radiation fields. In this configuration, the chamber without any deposited registered on-line the background signal. Cables were chosen to prevent leakage current higher than a few nA at 500 °C temperature. The bottom pair of fission chambers is shielded with $^{nat}$Gd filter to absorb thermal neutrons, i.e. become more sensitive to epithermal neutrons. Finally, one pair is constituted by a chamber with $^{241}$Am and one with $^{237}$Np. These different configurations are chosen to provide an overall characterisation of the inside neutron flux, in terms of its intensity but also its energy distribution. To increase the accuracy on the energy spectrum determination, nine activation neutron flux monitors were put inside the detector in a small titanium box. These monitors will be extracted and analysed during the post-irradiation phase.

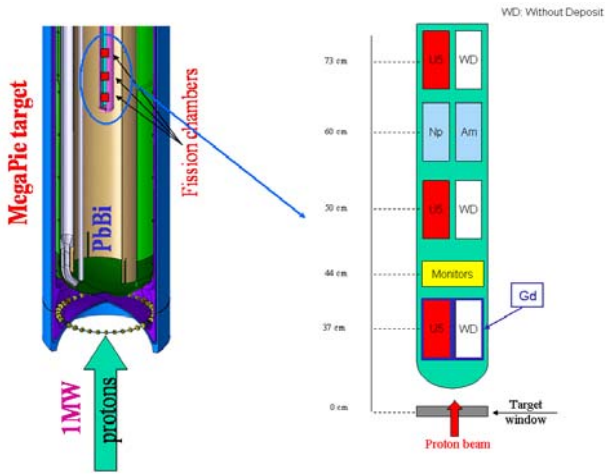

Fig. 2. Schematic view of the bottom part of the MEGAPIE target (left) and neutron detector layout (right).

## II.B. Delayed neutron (DN) flux measurement

Although the inner rod of MEGAPIE target was instrumented to measure precisely the PN flux, the DN flux was too small to be obtained with reasonable sensitivity by the same detector. Thus we decided to make use of another setup based on a $^3$He counter.

The DN detection was performed by a 45 cm long $^3$He (8 bar) tube installed in a polyethylene box (45x20x10 cm$^3$). The CH$_2$ box ensured the moderation of neutrons in order to increase the neutron detection efficiency. The CH$_2$ box was surrounded by a 1 mm thick $^{nat}$Cd foil to filter the room background of thermal neutrons. The detector was placed in the target head enclosure chamber, the so-called TKE, at around 3 m from the target head (Fig. 1).

The detector was set up and tested in the TKE at the end of June 2006. During this phase we performed a complete characterization of the detection system by using an Am-Be neutron source placed in different positions on the polyethylene box. The results of these tests have been compared to a simple MCNP simulation of the detector in order to estimate the neutron detection efficiency. This simulation showed a good agreement with the measured counting rate. However, in order to estimate the detector efficiency during the experiment, we still have to perform a complete MC simulation of the detector and its environment, taking into account that the DNs come from the whole LBE loop.

The DN data taking took place during the first week of the target irradiation. During this start-up phase, the beam power was increased gradually in long steps, each step being followed by a beam stop. This procedure gave us the possibility to acquire data at each beam stop, corresponding to variable beam powers.

## III. MODELLING OF THE TARGET

### III.A. Simulation of the prompt neutron flux

The MEGAPIE target is a complex system that has been simulated using Monte Carlo transport codes such as FLUKA[4] and MCNPX[5]. The first set of simulations[6] was performed in the target R&D phase in order to define the key parameters of the experiment (neutron flux intensity, mass of the deposit in the fission chambers, thermal/fast neutrons ratio, activation, etc…). The simulation work has been constantly improved, taking into account more and more detailed description of the whole SINQ geometry, including the target head, the TKE environment and the neutron guides[7]. The simulated neutron energy distributions are shown on Fig. 3. We can clearly see the effect of the Gd shielding on the lowest chambers where the thermal part of the spectrum is completely suppressed.

Since fission chambers are placed very close to the beam interaction point, they are very sensitive to the neutron flux distribution, both in position and energy. Thus it is very important to study the influence of the simulation parameters and physics models on the neutron flux properties. In particular, we performed a set of simulations using the MCNPX transport code to study the influence of the neutron detector geometry, the composition of the LBE, the beam profile and different spallation models on the neutron flux.

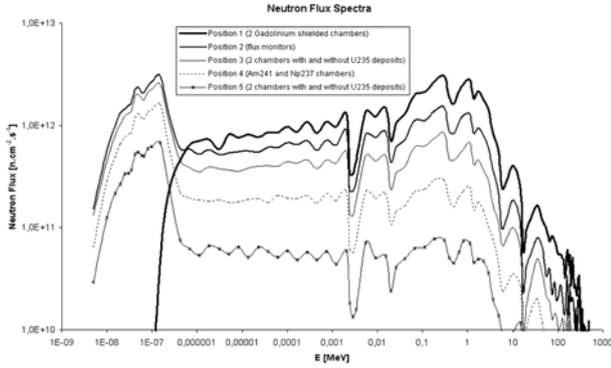

Fig. 3. Simulated neutron fluxes as "seen" by fission chambers, along the beam axis. Positions 1 and 5 correspond to the bottom and top chambers respectively.

First of all, our simulation shows that the presence of the neutron detector, described in details taking into account the presence of titanium pieces, filling gas and activation foils near the fission chambers, affects the neutron flux at the fission chambers positions not more than 1%.

The composition of the liquid Pb-Bi is a very important item because the presence of neutron absorbers (like boron, gadolinium or cadmium) in the LBE can modify significantly the energy distribution of the neutron flux including its absolute values. In order to compare the simulation to the quantity measured by the fission chambers, we present in Table I the simulated fission rate ($\sigma\phi$), which depends on the neutron flux and the effective fission cross section of the fissile isotope for the four positions along the neutron detector and for different concentrations of boron in the LBE. In particular, the fission rate is calculated taking the neutron flux distribution given by the MCNPX transport code and the fission cross sections from the ENDFB-VI library. We call "Std boron" the boron concentration which has been effectively measured from a sample of MEGAPIE LBE ($^{10}$B: 6.51 ppm; $^{11}$B: 27.2 ppm). This "real" composition is compared to the one without boron and two others with the "Std boron" concentration multiplied respectively by 10 and 100. The comparison shows that the presence of some ppm of boron in the LBE affects the neutron flux mostly in the thermal region, as expected. In the lowest position, where the Gd shielding cuts away the thermal part of the spectrum, the effect of B is around 3% and does not exceed 20% when the boron concentration reaches a few per mil. On the contrary, looking at the upper position, which is characterized by an almost fully moderated spectrum, the presence of some per mil of boron changes the fission rate by a factor of 3.

TABLE I. Simulated fission rates (per incident proton and per second) for different boron concentrations in LBE. The spallation model used is the MCNPX default one.

| Position (isotope) | 1 ($^{235}$U + Gd) | 3 ($^{235}$U) | 4 ($^{241}$Am) | 5 ($^{235}$U) |
|---|---|---|---|---|
| No boron | 6.74 e-10 | 8.63 e-9 | 3.92 e-11 | 3.08 e-9 |
| Std boron | 6.63 e-10 | 8.48 e-9 | 3.99 e-11 | 3.07 e-9 |
| Std x 10 | 6.43 e-10 | 7.28 e-9 | 3.66 e-11 | 2.83 e-9 |
| Std x 100 | 5.44 e-10 | 2.93 e-9 | 1.63 e-11 | 1.02 e-9 |

Another important parameter that can influence the neutron flux is the beam profile. In particular, since the fission chambers are placed in the beam axis and close to the impact point, the size and shape of the beam footprint can have a large impact on the measured flux. There exist different parameterizations of the beam footprint coming either from calculations or gamma activity measurements (made with different targets). Our study shows that the largest influence of the beam profile on the fission rate (around 14%) concerns mostly the lowest chambers, which are closer to the beam impact point. On the contrary, this effect does not exceed 4% for the upper chambers.

The last important item concerning the simulation of neutron flux is the evaluation of the influence of spallation models in the neutron production. The code MCNPX allows the user to choose between different intra-nuclear cascade and fission-evaporation model combinations among ISABEL, BERTINI and INCL4 for cascade and DRESNER and ABLA for de-excitation. The latest possibility with MCNPX is to use the package CEM2k (cascade and de-excitation). For both ISABEL and BERTINI models, the pre-equilibrium option has been used. Table II shows the simulated fission rate for different model combinations. From the simulated values we can see that the effect is not large, below 9%, but it should be looked at with care when one wants to perform precise studies on the neutron production. This is because the difference between models is more important in the epithermal and fast region of the neutron spectrum, while the effect becomes negligible in the thermal region.

Finally, it should be stressed that our present simulation does not take into account the actual temperature of Pb-Bi and D$_2$O moderator, which might have a non negligible influence. This is the main task of the ongoing simulation studies.

TABLE II. Simulated fission rates (per incident proton and per second) for different physics models within MCNPX. The simulation is performed with "std boron" concentration.

| Position (isotope) | 1 ($^{235}$U + Gd) | 3 ($^{235}$U) | 4 ($^{241}$Am) | 5 ($^{235}$U) |
|---|---|---|---|---|
| Bertini-Dresner | 6.63 e-10 | 8.48 e-9 | 3.99 e-11 | 3.07 e-9 |
| INCL4-ABLA | 7.01 e-10 | 8.29 e-9 | 3.83 e-11 | 3.07 e-9 |
| ISABEL-ABLA | 6.74 e-10 | 8.30 e-9 | 4.19 e-11 | 3.11 e-9 |
| CEM2k | 5.79 e-10 | 8.42 e-9 | 4.20 e-11 | 3.29 e-9 |

### III.B. Simulation of the delayed neutron (DN) flux

Liquid Pb-Bi eutectics (LBE) loop in the case of the MEGAPIE spallation target, as in most of the high power spallation targets based on liquid metal technologies, extends much further compared to the primary proton interaction zone. As it is presented in Fig. 4, the activated LBE has been transported over 400 cm when arriving in the heat exchanger, from where it returns to its initial position. It takes ~20 s for the entire ~82 liters of Pb-Bi to make a "round trip" at a flow rate of ~4 liters/s. It is clear that a big part of the DN precursors, created in the interaction region via high energy fission-spallation reactions, will not have enough time to decay completely even at the very top location of the loop. The main concern is about the DNs flux contributing to the total neutron flux at the very top position of the heat exchanger.

To estimate the DN flux we employed the multi-particle transport code MCNPX combined with the material evolution program CINDER'90 (Ref. 8), as detailed in Ref. 9. The DN data (emission probabilities and decay constants) were based on the ENDF/B-VI evaluations[10]. For the MEGAPIE target characteristics we used the design values, i.e. a 575 MeV proton beam with 1.75 mA intensity, interacting with the liquid LBE target. The 3-D geometry of the target has been modeled in detail by taking into account all materials used in the design, as described in Ref. 11.

The estimation of the DN parameters for MEGAPIE was performed in steps according to the following procedure:

− calculation of independent fission fragment and spallation product distributions with MCNPX
− calculation of cumulative fission fragment and spallation product yields with CINDER'90
− identification of all known DN precursors and construction of the 6-group DN table.

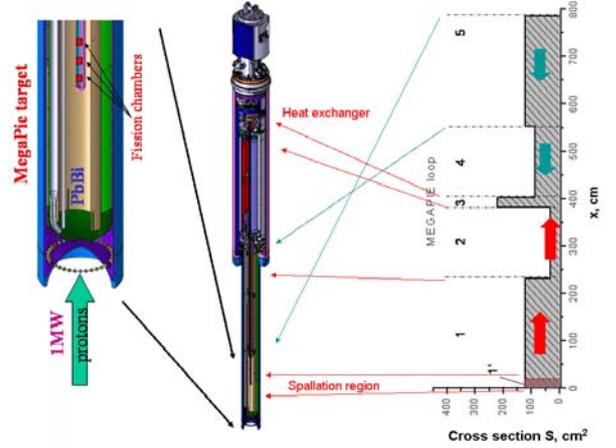

Fig. 4. Schematic view of the entire MEGAPIE target (in the middle) with a zoom of the lowest part – proton - Pb-Bi interaction zone (on the left). On the right: cross section $(cm^2)$ of the liquid LBE loop as a function of the Pb-Bi geometrical position – trajectory $x$ (cm).

After having built the DN table we developed a generalized geometrical model to estimate the DN activity densities at any position $x$ of the MEGAPIE target loop as presented in Fig 4. From this figure we can notice that the LBE cross section changes over the loop, meaning that the transit time of a given LBE volume depends on its position. In particular, the permanence time of the LBE volume under irradiation (in the so called spallation region) is very short (~0.5 s) compared to the total circulation time (~20 s). Within this model the DN density at position $x$ can be expressed as:

$$a(x) = \sum_{i=1}^{n} a_i(x) = \sum_{i=1}^{n} a_i \frac{1-\exp(-\lambda_i \tau_a)}{1-\exp(-\lambda_i T)} \exp(-\lambda_i \tau_d(x)), \quad (1)$$

where $\tau_a$ is the activation time of the Pb-Bi under irradiation; $T$ represents the total circulation period of the LBE, i.e. duration of the "round trip"; $\tau_d$ is the transit (decay) time to reach the point $x$; $\lambda_i$ is the decay constant of the DN precursor $i$, while $a_i$ stands for the density of DNs due to the precursor $i$. This equation is valid only at equilibrium, i.e. the irradiation time is large enough compared to DN precursor half-lives.

By the use of the above equation and the 6-group DN table[9] we found that at the very top position of the LBE loop (400 cm above the target window) the DN density is of the order of $2 \times 10^5$ n/(s cm$^3$). This intermediate result permitted to recalculate the neutron flux at the level of the heat exchanger inserting the volumetric DN source as a function of $x$ provided in Fig. 4. It was found that the neutron fluxes at this position due to DNs and prompt spallation neutrons are of the same order of magnitude, both equal to a few $10^6$ n/(s cm$^2$). It should be pointed out that this estimation rely on the hypothesis that 3 averaged time parameters are sufficient to describe a simplified liquid metal loop dynamics. These time constants, estimated from the target characteristics (LBE volume and main pump speed) are: *$\tau_a$=0.5s, T=20s* and *$\tau_d$ (at the heat exchanger)=10s* (see Eq. (1)).

In addition, prompt neutron energy spectrum at the heat exchanger position is very close to thermal (because the MEGAPIE spallation target is surrounded by a heavy water moderator-reflector) while the DN energy spectrum at this level is not yet moderated, i.e. with an average energy of the order of 400-600 keV. These fast neutrons will have considerably higher penetration power compared to the thermal ones. This result clearly points out that activation and dose rates due to DNs should not be neglected.

On the other hand, the 6-group DN parameters, i. e. yields and decay constant of DN precursors, which were extracted from MCNPX simulations based on different spallation models (namely INCL4+ABLA and CEM2k), are model-dependent nearly by two orders of magnitude[9]. This analysis showed that DN yields and time spectra from high energy fission-spallation reactions needed to be measured since no data of this type were available.

## IV. PRELIMINARY EXPERIMENTAL RESULTS

The MEGAPIE target has withstood for four months under a proton beam power close to 700 kW, instead of 1 MW, due to the difficulty to provide a stable proton beam over 1.2 mA. We present here a preliminary analysis on data taken during the first week of the target irradiation. During this start-up phase, the beam power was increased in long steps, giving the possibility to study the response of the PN detector and to acquire data for DN at each beam stop. During the whole irradiation period, only the PN measurement was performed.

### IV.A. Results from PN detector

The neutron detector has functioned reliably the whole irradiation phase at a temperature around 400°C with frequent beam interruptions. During this time the currents of the 8 fission chambers have been recorded every 2s. The beam current intensity was also recorded to study the neutron production of the target normalised to one incident proton, which is one of the fundamental items in the economy of an ADS neutron source.

The current measured by fission chambers is proportional to the fission rate ($\sigma\phi$) which depends on the neutron flux and the effective fission cross section of the fissile isotope. The extraction of the neutron flux is then not straightforward and depends on a good characterisation of the neutron energy distribution which is calculated with simulation codes. However, if the epithermal/thermal ratio does not evolve over time or with the beam intensity, the fission rate is a good estimate of the relative variations of the neutron flux. The evolution of the fission rate as a function of the proton beam intensity is shown on Fig. 5, where we see a good correlation for the three uranium chambers, as expected. Note that at this stage the burn-up can be considered as negligible. This validates the correct functioning of the detectors. On Fig. 6, the evolution of the fission rate normalised to the proton beam intensity is plotted, as a function of time for the whole irradiation period, for the middle and the upper chambers. We can see a small decrease of the fission currents due to the burn-up of the uranium deposit, estimated around 6%.

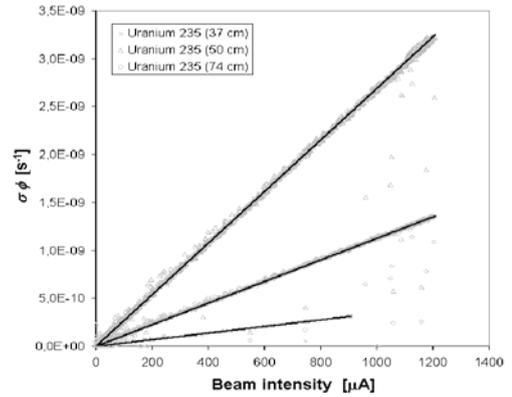

Fig. 5. Correlation between the fission rate and the proton beam intensity.

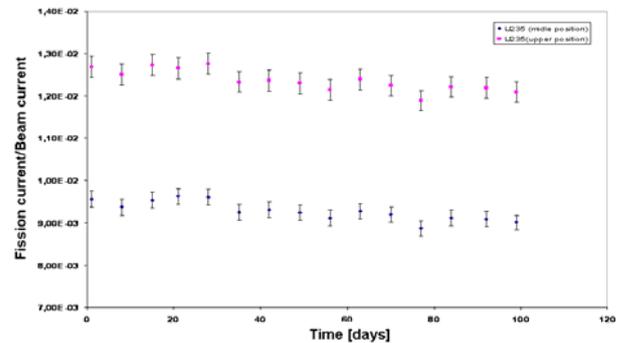

Fig. 6. Time evolution of two $^{235}$U fission chambers, normalised to the proton beam current.

Taking into account all improvements and studies of the target system description, as described in the previous section, we compared the measured fission rate to the simulated one (Table III). The fission rate normalization is actually extracted from the measured fission current taking into account the chamber sensitivity which has been previously measured at ILL with a precision of 3%.

TABLE III. Comparison of simulated fission rates (per incident proton and per second) with the measured ones for different fission chambers.

| Position (isotope) | 1 ($^{235}$U + Gd) | 2 ($^{235}$U) | 3 ($^{241}$Am) | 4 ($^{235}$U) |
|---|---|---|---|---|
| Measured $\sigma\phi$ (uncertainty) | 3.70E-10 (3%) | 2.85E-9 (3%) | 1.37E-11 (3%) | 1.19E-9 (3%) |
| Simulated $\sigma\phi$ | 6.63 e-10 | 8.48 e-9 | 3.99 e-11 | 3.07 e-9 |

From this comparison we can see that there is a systematic over-prediction of the measured values by a factor of 2-3, which cannot be explained for the moment. Our work continues concentrating in both simulation and data analysis to disentangle all the different contributions affecting the neutron flux simulation (temperature, target movement…) and the fission rate measurement (sensitivity, burn-up…).

### IV.B. Results from DN detector

We recall that the DN data taking took place during the first week of the target irradiation in order to acquire data at each beam stop, corresponding to variable beam powers.

The measured counting rate as a function of time, normalized to the beam intensity, is presented on Fig. 7 for different beam powers. We can notice that during the irradiation the neutron detector saturates, due to the high neutron flux, i.e. the detector gives a counting rate independent from the beam power. When the beam stops, we start counting DNs but some seconds are needed before the counting rate becomes proportional to the beam intensity. Since we know that the detector electronics needs ~50 ms after saturation to become operational again, we can argue that the DN flux during the first seconds after the beam stop is still quite high (from calculations of the same order of PN flux) and the detector is still in saturation. As soon as the DN flux lowers, the counting rate becomes proportional to the beam intensity: as expected, at equilibrium the DN precursor production rates are proportional to the beam power. Thanks to this proportionality, we can sum the decay curves taken at different beam power to increase the statistics.

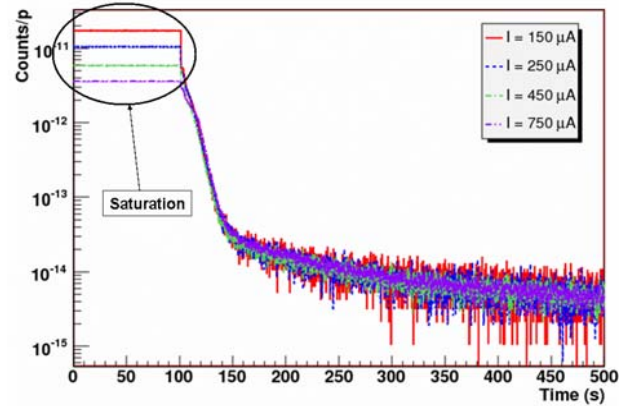

Fig. 7. The DN decay curves normalized to the beam intensity; different curves represent irradiations at different beam powers (see the legend).

The interpretation of the data is not trivial since we do not know "a priori" which precursors are contributing to the DN flux. On the other hand, DNs were measured using 1 GeV protons interacting with massive Pb and Bi targets of variable thicknesses at PNPI Gatchina (Russia)[12]. During this experiment it was found that, contrary to the conventional 6-group approach, DN decay curves could be described by 4 exponential terms corresponding to four dominant isotopes. In particular, up to 10-20 s, major DN contributors come from light mass products, resulting from the spallation process, as $^9$Li and $^{17}$N, rather than fission products as in the case of actinide fission. For longer decay time, from 50 to 100 s, the DN activity is dominated by usual fission products as $^{88}$Br and $^{87}$Br. This result has been a starting point for the MEGAPIE data interpretation.

Using the normalized decay time spectra shown in Fig. 7, summed up to increase the statistics, and taking the DN precursor half-lives extracted from the experiment at PNPI[12], we fit the experimental decay curve using Eq. (1). From the fit (Fig. 8) we extracted the DN densities $a_i$ (normalised to unity) of the identified precursors, together with the LBE transit times $\tau_a$ and $T$.

Table IV summarizes the results obtained from the fit. We add that, as for the Gatchina experiment, it was impossible to extract DN density for $^9$Li due to its short half-life, which is comparable with the acquisition channel width $\Delta t_{ch}$ = 200 ms. Since the analysis is still ongoing and the results are preliminary, we decided not to quote error bars on DN densities.

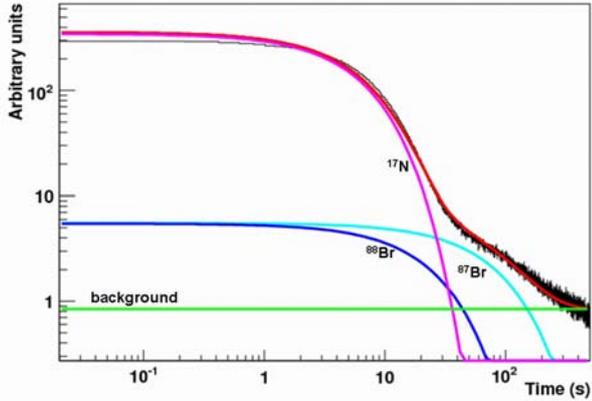

Fig. 8. Fit of the experimental DN decay curve obtained from Eq. (1). The relative contributions from individual precursors are also shown. The zero on the time scale corresponds to the end of the proton pulse. Note the arbitrary units in the figure.

TABLE IV. DN densities $a_i$ (normalised to unity) of the 3 identified contributors extracted from experimental fit.

| Group | Precursor | Half-life[11], s | $a_i$, % |
|---|---|---|---|
| 1 | $^{87}$Br | 55.60 | 4.3 |
| 2 | $^{88}$Br | 16.29 | 3.3 |
| 3 | $^{17}$N | 4.173 | 92.4 |

The DN densities are fairly compatible with Gatchina experiment results, meaning that in the LBE loop the precursors involved are the same as in the solid target experiment. Moreover, the LBE transit times from the fit ($\tau_a$=0.49 s, T=19.6 s) are in very good agreement with the values estimated from the loop technical characteristics. This means that the LBE loop can be well approximated by the three averaged time parameters, which validates the simplified approach developed above.

We should conclude by stressing that these preliminary results are not an absolute measurement of DN flux since a detailed simulation of the detector efficiency is still ongoing.

## V. CONCLUSIONS

We presented a set of studies on the neutronics of the MEGAPIE spallation target, based on simulations, modelling and measurements of PN and DN flux. As a general comment, we want to point out that all the simulation studies show the importance of the implantation of the neutron detector inside the target to study macroscopic effects that could greatly modify estimated quantities as, for example, activation residues. Moreover, a preliminary fit of the DN decay curve measured at MEGAPIE was performed using the result of a geometrical model involving three averaged liquid metal transit times. The DN densities extracted are in fair agreement with DN parameters previously measured at similar energy but with solid targets in simple geometrical configuration.

## ACKNOWLEDGMENTS

The authors are grateful to the MEGAPIE project management and PSI-SINQ staff, in particular Luca Zanini, for cooperation and assistance during the detector installation and data taking. This work was partially supported by the MEGAPIE initiative and also by the GEDEPEON collaboration (France).